\documentstyle[aps,prd,eqsecnum,twocolumn,epsf]{revtex}
\draft

\begin{document}

\draft
\twocolumn[\hsize\textwidth\columnwidth\hsize\csname
@twocolumnfalse\endcsname

% For two column
%\wideabs{

\title{Density Discontinuity of a Neutron Star and Gravitational Waves}

\author{
Hajime Sotani$^{1}$
\thanks{Electronic address:sotani@gravity.phys.waseda.ac.jp},
Kazuhiro Tominaga$^{1}$
\thanks{Electronic address:tominaga@gravity.phys.waseda.ac.jp}, and
Kei-ichi Maeda$^{1,2}$
\thanks{Electronic address:maeda@gravity.phys.waseda.ac.jp}
}

\address{
$^{1}$ Department of Physics, Waseda University,
3-4-1 Okubo, Shinjuku, Tokyo 169-8555, Japan
}

\address{
$^{2}$ Advanced Research Institute for Science and Engineering, \\
Waseda University, Shinjuku, Tokyo 169-8555, Japan
}

\date{\today}

\maketitle

% Abstract
\begin{abstract}
We calculate quasi-normal f- and g-modes of a neutron star with density
discontinuity,
which may appear in a phase transition at extreme high density. We find
that  discontinuity will reflect largely on
  the f-mode, and  that the  g-mode could also be important
for a less massive  star.
\end{abstract}


]

% For two column
%}

%\baselineskip 24pt

% Introduction
\section{Introduction}

Gravitational wave interferometers such as LIGO, VIRGO, GEO600,
and TAMA300 on earth, and LISA in space \cite{Thorne}
are almost ready to observe directly the gravitational
waves from the Universe.
The significance of direct observation of gravitational waves is that we obtain
  new method to observe the Universe  by gravitational waves (the so-called
gravitational wave astronomy),  and that we may be able to  verify the
theory of
gravity, and that new physics at  high density or high energy region can be
found by
observing a merger of a binary neutron star.

The one of the sources of the gravitational waves is supposed to be
supernova explosion,
which occurs at the last phase of a massive star. After explosion, a newly
formed
compact  object may oscillate  violently, and the emitted gravitational
waves carry  the information of the sources. The oscillation will
eventually damp out
because that the gravitational waves  carry the
energy. The merger of two coalescing black holes or
neutron stars is another  source. It is naturally expected that the final
object will also oscillate.
  If a final
compact object is a neutron star, its structure will reflect on the emitted
gravitational
waves. There should be some  relation between
a structure of  a neutron star and the emitted  gravitational waves.
The structure
of a neutron star not yet well-known because the equation of  state is
uncertain at the high density over the nuclear density.
Therefore, although main
composition of matter  in a neutron star is neutron,  it is possible that
there appear
new phase in extremely high density such as
a pion condensation, or a kaon  condensation.  In such a  situation,
a phase transition may occur, and if it is strongly first order, one may
expect that there
appears the discontinuity of density at some density as the case of the
kaon condensation
\cite{Tatsumi}.
If we  know in advance the relation between the structure of a neutron star
and the emitted gravitational waves, we will  get the information or the
restriction on the equation of state over the nuclear density.
In particular, if we find any specific feature in the emitted gravitational
waves from a
neutron star with discontinuity, we may be able to predict the existence of
a phase
transition in a neutron star.
This is the main subject of the present paper.
We will just focus into quasi-normal modes of  a neutron star.

Although the gravitational waves emitted in a supernova explosion or from a
merger of a
binary may show very complicated wave form, only  quasi-normal modes will
eventually
remain.  We expect that these quasi normal modes contain a lot of
information about a
neutron star. Hence, it is very important for us to study those modes for a
possible
situation.
The quasi normal modes are classified into
various types. For the modes coupled to the fluid of
a non-rotating neutron star, we have  f-, p- and g-modes.

The damping
rate of these modes is very small  compared with the oscillation period of
fluid
because  matter couples to gravity very weakly.
There exists only one f-mode for each $l$.
The  p-mode which is caused by the  pressure of fluid.
The g-mode is originated by density discontinuity or temperature of a star.
Its damping
rate is usually quite small, so  few calculation has been done.  There also
exists
another mode (the so-called w-mode) which is related to metric perturbations of
spacetime  rather than  fluid oscillation. This mode has high damping rate
in comparison with  modes related to fluid oscillation.

For the case without density discontinuity, the
relation between quasi normal modes of a star and
equation of state has been studied by many authors\cite{Andersson1996},
\cite{Andersson1998},
\cite{Andersson1999}. For g-mode, there are a few works. Newtonian case is
studied in \cite{McDermott}  and also in \cite{Strohmayer}, which includes
the effect of
temperature of neutron stars by use of the Cowling
approximation\cite{Cowling}.
For the relativistic case, Finn adopted an approximate method which is
called the ``slow
motion formalism" and calculated the imaginary part  by the energy loss for the
case of density discontinuity \cite{Finn1}, \cite{Finn2}.

Although these works are very important, they have not discussed density
discontinuity
at high density, which may appear via a phase transition.
  In this paper, we shall study a neutron star with such  a discontinuity,
and investigate
its quasi normal modes.
Just for  simplicity, we assume that
equation of  state is given by a polytrope ($P=K\rho^{1+1/n}$).
We discuss only  even parity modes, which are coupled to oscillation of
matter fluid in a  neutron star.
We then assume that there  exists density discontinuity due to a phase
transition and density discontinuity.

First we present the basic equations to solve in \S
\ref{sec:basic_equation}, including the equation for a background neutron
star with
density discontinuity.
  In \S \ref{sec:numerical_results}, we
show an explicit structure of a neutron star  with the
density discontinuity
  and  numerical results for quasi normal modes. We also compare our results
with those
obtained by the  relativistic Cowling approximation\cite{McDermott1983}.
Conclusion
is given in \S\ref{sec:conclusion}. In Appendix
\ref{sec:appendix_1}, we present the explicit boundary condition at the
center of
a star. In Appendix \ref{sec:appendix_2}, we briefly summarize the method
of how to calculate the quasi normal modes.

In this paper we adopt the unit of $c=G=1$, where $c$ and $G$ denote the
speed of light and the gravitational constant, respectively, and the metric
signature of $(-,+,+,+)$.

\section{basic equation}
\label{sec:basic_equation}

\subsection{stellar model with discontinuity}

We first construct a neutron star model with discontinuity of density to
calculate
its quasi-normal modes. A static and
spherically symmetric spacetime is described by
  \begin{equation}
   ds^{2}=-e^{2\Phi}dt^{2}+e^{2\Lambda}dr^{2}+r^{2}\left(d\theta^{2}+\sin^{2}
          \theta d\phi^{2}\right),
\label{metric_star}
  \end{equation}
where $\Phi,\Lambda$ are metric functions with respect to $r$. A  mass function
$m(r)$, which is defined as
  \begin{equation}
   m(r) = \frac{1}{2}r\left(1-e^{-2\Lambda}\right), \label{ephi}
  \end{equation}
satisfies
\begin{equation}
  \frac{dm}{dr}    = 4\pi r^2\rho . \label{dmdr}
  \end{equation}
The equilibrium condition of a stellar model is given by the
Tolman-Oppenheimer-Volkoff equation,
  \begin{equation}
   \frac{dP}{dr}   = -\frac{\left(\rho+P\right)\left(m+4\pi r^3P\right)}
                        {r\left(r-2m\right)}, \label{dpdr}
  \end{equation}
where the function $\rho$ and $P$  are the energy
density and the pressure of the fluid.
The potential $\Phi$ is given by
  \begin{equation}
   \frac{d\Phi}{dr}   = \frac{\left(m+4\pi r^3P\right)}
                        {r\left(r-2m\right)}. \label{dphidr}
  \end{equation}
With an additional equation, i.e. the equation of state, we can  solve
these equations.

As for the boundary conditions, there are two boundaries; the center of a
star and the
surface.
The stellar radius
$R$ is determined by the condition that the pressure vanishes at the
surface, i.e.
$P(R)=0$. The total mass $M$ is given by $m(R)$, and the potential
$\Phi(r)$ is smoothly
connected at the surface, i.e.
  \begin{equation}
   e^{2\Phi(R)} = 1-\frac{2M}{R}.
  \end{equation}
At the center, a regularity condition should be imposed.

As for the equation of state, just for simplicity, we adopt a polytropic
equation of
state except at a  discontinuous surface of  density, which is assumed to
exist inside
of the star. Then, the equation of state is given by two equations;
  \begin{eqnarray}
   P = K_{(+)}\rho^{1+1/n_{(+)}} &\hspace{5mm} \mbox{for} \hspace{2mm}
                0\le r\le R_{\rm dis}, \\
   P = K_{(-)}\rho^{1+1/n_{(-)}} &\hspace{5mm} \mbox{for} \hspace{2mm}
                R_{\rm dis}\le r\le R,
  \end{eqnarray}
where $R_{\rm dis}$ is the radius of a discontinuous surface.
The values of two adiabatic indecies ($n_{(+)}$ and $n_{(-)}$), and those
of two
coefficients  ($K_{(+)}$ and $K_{(-)}$) may be different.
However, since the pressure should be continuos at the discontinuous
surface $R_{\rm dis}$, we have one constraint on those parameters. For
example,
  $K_{(+)}$ is fixed as
  \begin{equation}
   K_{(+)} = K_{(-)}\times
    \frac{\left(\rho_{\rm
dis}^{(-)}\right)^{[1+1/n_{(-)}]}}{\left(\rho_{\rm
dis}^{(+)}\right)^{[1+1/n_{(+)}]}}.
  \end{equation}

\begin{center}
\begin{figure}[htbp]
\epsfxsize = 3in
\epsffile{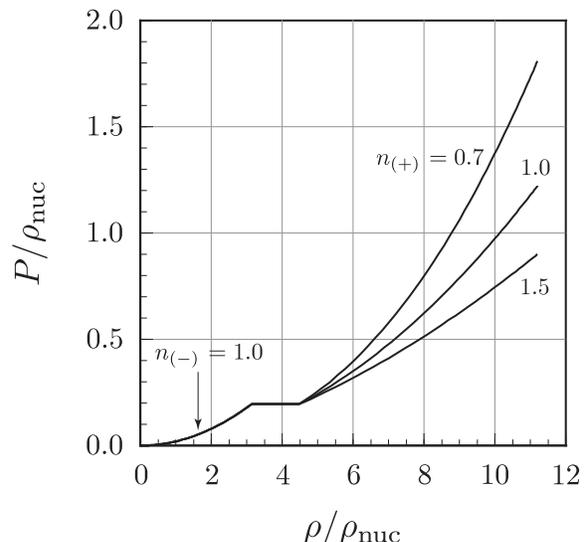}
\caption{Equation of state with density discontinuity. The density and
pressure are normalized by  the standard nuclear density $\rho_{\rm
nuc}=2.68
\times 10^{14}$\,(g/cm$^3$).
We set  $\rho_c=3.0 \times 10^{15}$\,(g/cm$^3$),
$\rho_{\rm
dis}^{(+)}=1.2 \times 10^{15}$\,(g/cm$^3$), $\rho_{\rm
dis}^{(-)}/\rho_{\rm
dis}^{(+)}=0.7$,
$K_{(-)}=100$\,(km$^2$) and $n_{(-)}=1.0$.}
\label{fig_eos}
\end{figure}
\end{center}

Now we have six parameters;
the central density of the star $\rho_c$, the inner density at the
discontinuous surface $\rho_{\rm
dis}^{(+)}$, the ratio of the outer density
to the inner density at the discontinuous surface
$\rho_{\rm
dis}^{(-)}/\rho_{\rm
dis}^{(+)}$, the inner polytropic index $n_{(+)}$,
the outer polytropic index $n_{(-)}$ and the outer polytropic coefficient
$K_{(-)}$.
Some examples of this equation of state with discontinuous density
are given in Figure \ref{fig_eos}.

% Figure 1

\onecolumn

\subsection{perturbation equations inside the star}

Here we present the basic equations for perturbations against the
background metric of
a spherically symmetric star (\ref{metric_star}).  The perturbation is given by
  \begin{equation}
   g_{\mu\nu} = g^{(B)}_{\mu\nu}+h_{\mu\nu}.
  \end{equation}
The perturbations against a spherical background can be
decomposed by the spherical harmonics $Y^{l}_{m}$, and are classified into two
by its parity, i.e. odd or even.  Because we are interested in to  discuss
relation between a discontinuity in stellar  density and the  gravitational
waves
emitted from such a neutron star, in this paper, we
only study even-parity perturbations, in which gravitational waves are
coupled to fluid
oscillation.

Following the method by Lindblom and Detweiler\cite{LindblomDeweiler},
$h_{\mu\nu}$ is described as
  \begin{equation} 
   h_{\mu\nu} = \left( 
   \begin{array}{cccc} 
   r^l\hat{H}e^{2\Phi} & i\omega r^{l+1}\hat{H}_{1} & 0 & 0  \\
   i\omega r^{l+1}\hat{H}_{1} & r^l\hat{H}e^{2\Lambda} & 0 & 0  \\
   0 & 0 & r^{l+2}\hat{K} & 0 \\
   0 & 0 & 0              & r^{l+2}\hat{K}\sin^{2}\theta
  \end{array} 
  \right) Y^{l}_{m}\,e^{i\omega t},
  \end{equation}
where $\hat{H},\hat{H}_{1},$ and $\hat{K}$ are the perturbed metric
function with
respect to  $r$.
The fluid perturbations  are described
by the Lagrangian displacement vectors $\xi^{i} =(\xi^r, \xi^\theta,
\xi^\phi)$ as
follows:
  \begin{eqnarray}
   \xi^{r}      &=& \frac{r^l}{r}e^{\Lambda}\hat{W}Y^{l}_{m}\,
                    e^{i\omega t}, \\
   \xi^{\theta} &=& -\frac{r^l}{r^2}e^{\Lambda}\hat{V}
                    \frac{\partial}{\partial \theta}
                    Y^{l}_{m}\,e^{i\omega t}, \\
   \xi^{\phi}   &=& -\frac{r^l}{r^2\sin^2\theta}e^{\Lambda}\hat{V}
                    \frac{\partial}{\partial \phi}Y^{l}_{m}\,e^{i\omega t},
  \end{eqnarray}
where $\hat{W} $ and $ \hat{V}$ are  functions with respect to $r$.
Furthermore we introduce the new variable $\hat{X}$ instead of $\hat{V}$ to
make the boundary condition at the surface of a star simple (see below);
  \begin{equation}
   \hat{X} \equiv \omega^2\left(P+\rho\right)e^{-\Phi}\hat{V}
                  -\frac{e^{\Phi-\Lambda}}{r}\frac{dP}{dr}\hat{W}
                  -\frac{1}{2}\left(P+\rho\right)e^{\Phi}\hat{H}.
  \end{equation}
The perturbation equations of  the
Einstein equations are given by
  \begin{eqnarray}
   &&\frac{d\hat{H}_{1}}{dr}=-\frac{1}{r}
    \left[l+1+\frac{2m}{r}e^{2\Lambda}+4\pi r^2(P-\rho)e^{2\Lambda}\right]
    \hat{H}_{1} \nonumber \\
   &&\hspace{1.2cm}+\frac{1}{r}e^{2\Lambda}
    \left[\hat{H}+\hat{K}+16\pi (P+\rho)\hat{V}\right],\label{perturbation1} \\
   &&\frac{d\hat{K}}{dr}=\frac{l(l+1)}{2r}\hat{H}_{1}
    +\frac{1}{r}\hat{H}
    -\left(\frac{l+1}{r}-\frac{d\Phi}{dr}\right)\hat{K}
    +\frac{8\pi}{r}(P+\rho)e^{\Lambda}\hat{W}, \label{perturbation2} \\
   &&\frac{d\hat{W}}{dr}=-\frac{l+1}{r}\hat{W}+re^{\Lambda}
    \left[\frac{1}{\gamma P}e^{-\Phi}\hat{X}-\frac{l(l+1)}{r^{2}}\hat{V}
    -\frac{1}{2}\hat{H}-\hat{K}\right], \label{perturbation3} \\
   &&\frac{d\hat{X}}{dr}=-\frac{l}{r}\hat{X}+(P+\rho)e^{\Phi}
    \biggl[\frac{1}{2}\left(\frac{d\Phi}{dr}-\frac{1}{r}\right)\hat{H}
    -\frac{1}{2}\left(\omega^{2}re^{-2\Phi}+\frac{l(l+1)}{2r}\right)
    \hat{H}_{1}  \nonumber \\
   &&\hspace{1.2cm}+\left(\frac{1}{2r}-\frac{3}{2}\frac{d\Phi}{dr}\right)
    \hat{K}-\frac{l(l+1)}{r^{2}}\frac{d\Phi}{dr}\hat{V}
    \nonumber \\
   &&\hspace{1.2cm}-\frac{1}{r}
    \left(\omega^{2}e^{-2\Phi+\Lambda}+4\pi(P+\rho)e^{\Lambda}
    -r^{2}\left\{\frac{d}{dr}\left(\frac{1}{r^{2}}e^{-\Lambda}\frac{d\Phi}{dr}
    \right)\right\}\right)\hat{W}\biggr], \label{perturbation4} \\
   &&\biggl[1-\frac{3m}{r}-\frac{l(l+1)}{2}-4\pi r^{2}P\biggr]\hat{H}
    -8\pi r^{2}e^{-\Phi}\hat{X} \nonumber \\
   &&\hspace{1.2cm}+r^{2}e^{-2\Lambda}\left[\omega^{2}e^{-2\Phi}
    -\frac{l(l+1)}{2r}\frac{d\Phi}{dr}\right]\hat{H}_{1}
    \nonumber \\
   &&\hspace{1.2cm}-\left[1+\omega^{2}r^{2}e^{-2\Phi}-\frac{l(l+1)}{2}
    -\left(r-3m-4\pi r^{3}P\right)\frac{d\Phi}{dr}\right]\hat{K}=0,
    \label{perturbation5} \\
   &&\hat{V}=\frac{e^{2\Phi}}{\omega^{2}(P+\rho)}
    \left[e^{-\Phi}\hat{X}
    +\frac{1}{r}\frac{dP}{dr}e^{-\Lambda}\hat{W}+\frac{1}{2}(P+\rho)\hat{H}
    \right] ,
\label{perturbation6}
  \end{eqnarray}
where $\gamma$ is the adiabatic index of the unperturbed
stellar model defined by
  \begin{equation}
   \gamma = \frac{\rho+ P}{P}\left(\frac{\partial P}{\partial
\rho}\right)_{\rm ad}.
        \label{gamma}
  \end{equation}
Eqs. (\ref{perturbation1})$-$(\ref{perturbation4})
give a set of differential equations for the variables
$\hat{H}_1$, $\hat{K}$, $\hat{W}$ and $\hat{X}$, while Eqs.
(\ref{perturbation5}) and
(\ref{perturbation6})
are the algebraic equations for the  variables $\hat{H}$ and $\hat{V}$.

\subsection{the perturbations outside the stars}

In the region outside the star, the perturbations are  described by the
Zerilli equations. Setting  $m=M,\hat{W}=\hat{V}=0$,
and replacing  $\hat{H}_{1},\hat{K}$ with new variables
$Z$, $dZ/dr_{*}$ defined by
  \begin{eqnarray}
   r^{l}\hat{K} &=&
     \frac{\lambda(\lambda+1)r^{2}+3\lambda Mr+6M^{2}}{r^{2}(\lambda r+3M)}Z
     +\frac{dZ}{dr_{*}}, \label{khyoumenn} \\
   r^{l+1}\hat{H}_1 &=&\frac{\lambda r^{2}-3\lambda Mr-3M^{2}}
     {(r-2M)(\lambda r+3M)}Z+\frac{r^2}{r-2M}\frac{dZ}{dr_{*}},
     \label{hhyoumenn}
  \end{eqnarray}
we recover
the Zerilli equations.
Here  $r_{*}$ is the tortoise coordinates and $\lambda\equiv
l(l+1)/2-1$.

\subsection{boundary conditions}

In order to solve the above perturbation equations, we have to
  impose
the boundary conditions. At the center of the star, we have regularity
condition for the perturbative variables. We expand all variables by Taylor
power-series  near $r=0$ in a similar way of Detweiler and
Lindblom \cite{LindblomDeweiler} (see Appendix \ref{sec:appendix_1}).
Another boundary
condition is that  the pressure vanishes at the stellar surface. Then the
perturbed pressure
$\hat{X}$, $\Delta P$ is given by
  \begin{equation}
   \Delta P = -r^le^{-\Phi}\hat{X}
  \end{equation}
  also vanishes there. This boundary condition at the surface of star is
equivalent to  $\hat{X}(R)=0$.
The last condition is that there exist only outgoing gravitational waves
at infinity. We also have a junction condition at the
surface of   discontinuity, where the
variables of $\hat{H}_1,\hat{K},\hat{W},\Delta P$ (and then $\hat{X}$)
  must be continuous.
Note that, $\hat{H}$ and $\hat{V}$ are determined by Eqs. 
(\ref{perturbation5})  and (\ref{perturbation6}).
$\hat{H}$ is continuos but $\hat{V}$ is discontinuous.

\section{numerical results}
\label{sec:numerical_results}

\subsection{models of neutron star}

As a concrete example of density discontinuity, we know  kaon condensation
\cite{Tatsumi}, which may  occur at $\rho_{\rm
dis}^{(+)}=1.466 \times 10^{15}$\,g/cm$^3$ with the discontinuity
$\rho_{\rm
dis}^{(-)}/\rho_{\rm
dis}^{(+)}=0.414$ for a cold neutron star. Hence we consider density
discontinuity  around this value, i.e. we study  the following three cases:
$\rho_{\rm
dis}^{(+)}=8.0 \times 10^{14}$, $1.2 \times 10^{15}$ and
$1.6 \times 10^{15}$\,g/cm$^3$.
First  we briefly discuss a models of a neutron star
with density discontinuity.
As mentioned before, we have six unknown  parameters: $n_{(-)},
K_{(-)}, n_{(+)}, \rho_{\rm dis}^{(+)},  \rho_{\rm
dis}^{(-)}/\rho_{\rm dis}^{(+)}$
and $\rho_c$.
We set  $n_{(-)}=1.0$ and
$K_{(-)}=100\,(\text{km}^2)$, which are usually  adopted  in a neutron star
model.
The rest four  are
$n_{(+)}, \rho_{\rm
dis}^{(+)},
\rho_{\rm dis}^{(-)}/\rho_{\rm
dis}^{(+)}$
and $\rho_c$.

\begin{center}
\begin{figure}[htbp]
\epsfxsize = 6in
\epsffile{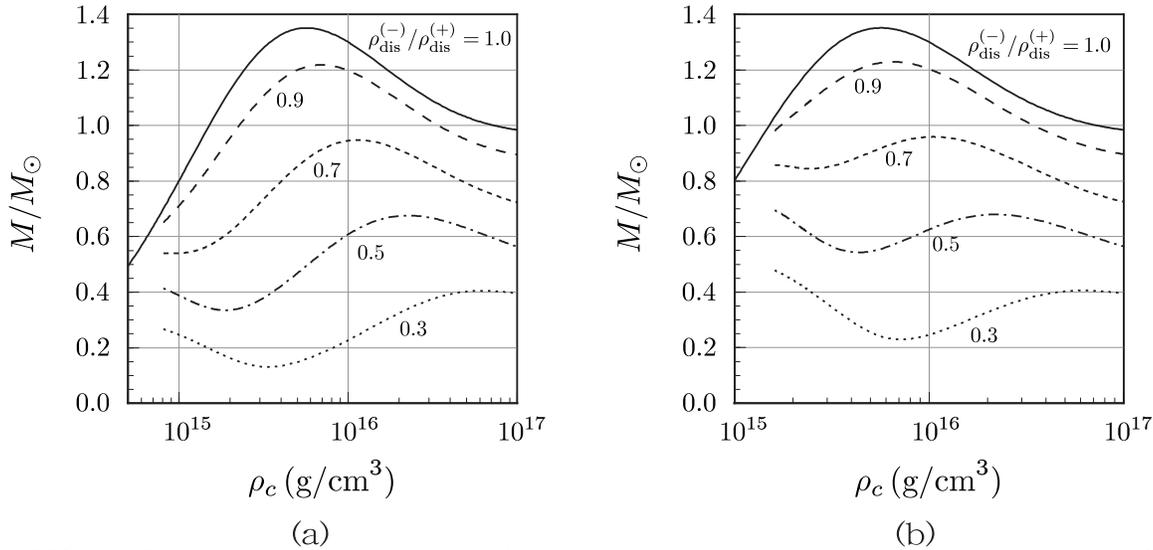}
\caption{
The relation between the central density $\rho_c$ and the gravitational
mass of a
neutron star $M$ for
$n_{(+)}=1.0$. Fig. 2(a) is the case of $\rho_{\rm
dis}^{(+)}=8.0 \times 10^{14}$
\,(g/cm$^3$) , while Fig. 2(b) is the case of $\rho_{\rm
dis}^{(+)}=1.6 \times 10^{15}$
\,(g/cm$^3$). The numbers associated with each curve are the values of
$\rho_{\rm
dis}^{(-)}/\rho_{\rm
dis}^{(+)}$.
}
\label{fig_rhoc_M_10}
\end{figure}
\end{center}

\begin{center}
\begin{figure}[htbp]
\epsfxsize = 6in
\epsffile{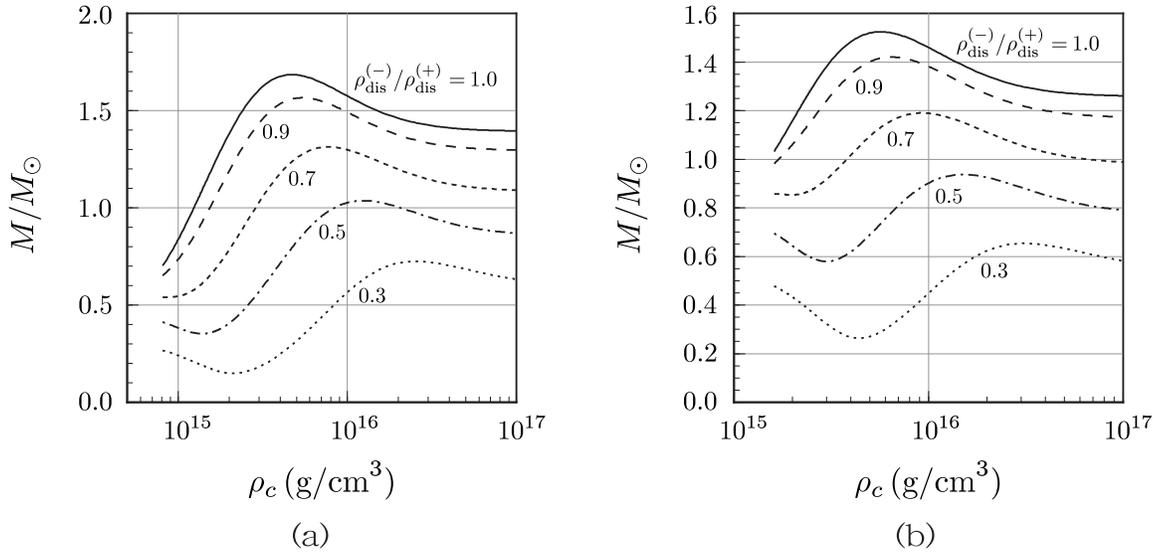}
\caption{
The same figures as Fig. 2  for the case of
$n_{(+)}=0.7$. }
\label{fig_rhoc_M_07}
\end{figure}
\end{center}

\begin{center}
\begin{figure}[htbp]
\epsfxsize = 6in
\epsffile{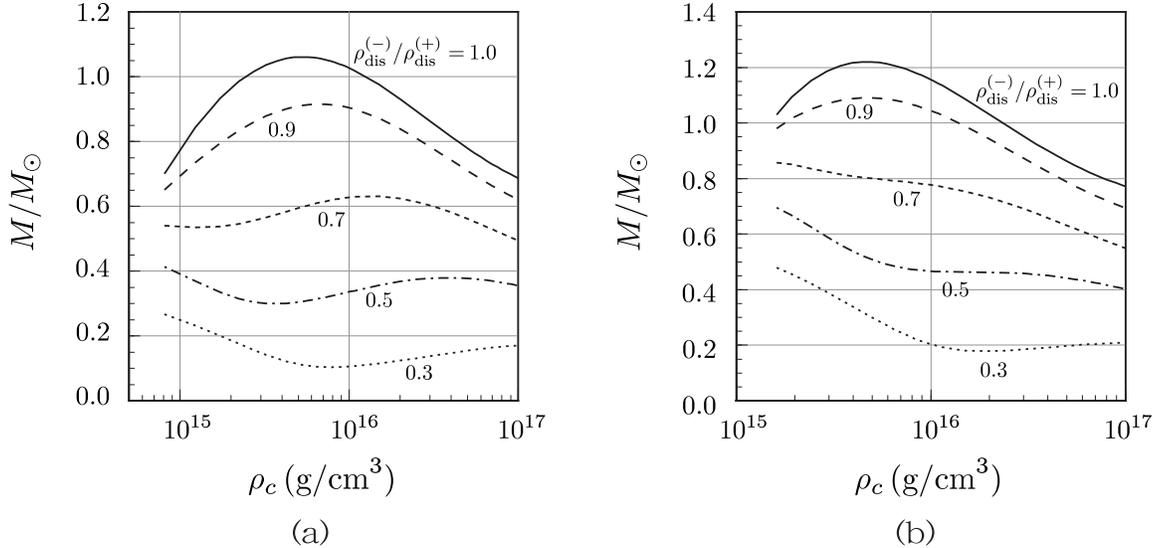}
\caption{
The same figures as Fig. 2 for the case of
$n_{(+)}=1.5$. }
\label{fig_rhoc_M_15}
\end{figure}
\end{center}

First we show the gravitational mass $M$ with respect to a central density
$\rho_c$
for several values of $n_{(+)}, \rho_{\rm
dis}^{(+)},
\rho_{\rm dis}^{(-)}/\rho_{\rm
dis}^{(+)}$ in Figs. \ref{fig_rhoc_M_10}-\ref{fig_rhoc_M_15}.

% Figure 2

% Figure 3

As for inner polytropic index, we
set  $n_{(+)}=$ 1.0 (Fig.
\ref{fig_rhoc_M_10}), 0.7 (Fig. \ref{fig_rhoc_M_07}), 1.5 (Fig.
\ref{fig_rhoc_M_15}).
For each polytropic index, we consider two models with different
density at discontinuity:$\rho_{\rm dis}^{(+)}=8.0 \times 10^{14}$ (a) and
$1.6 \times 10^{15}$\,g/cm$^3$ (b).
We depict the $M$-$\rho_c$ diagram for each value of  $\rho_{\rm
dis}^{(-)}/\rho_{\rm
dis}^{(+)}$ = 1.0 (no discontinuity), 0.9, 0.7, 0.5,  and 0.3.
Fig. \ref{fig_rhoc_M_10} for $n_{(+)}=1.0 $ shows that the maximum masses
for each value
of
$\rho_{\rm
dis}^{(-)}/\rho_{\rm
dis}^{(+)}$ do not depend on density discontinuity
$\rho_{\rm
dis}^{(+)}$. As $\rho_{\rm
dis}^{(-)}/\rho_{\rm
dis}^{(+)}$ gets small, the maximum mass decreases and the stable
region ($dM/d\rho_c>0$) becomes narrower  and moves to high
  density region.

For $n_{(+)}\neq 1.0$, the maximum mass depends on $\rho_{\rm dis}^{(+)}$.
In fact, for $n_{(+)}<1.0$ (Fig. \ref{fig_rhoc_M_07}), the maximum mass
decreases as
$\rho_{\rm dis}^{(+)}$ increases, while for $n_{(+)}>1.0$ (Fig.
\ref{fig_rhoc_M_15}),
the maximum mass increases.
As the equation of state of inner region becomes more stiff, i.e. as $n_{(+)}$
decreases,
the maximum mass gets large. The other properties about maximum mass   for
$n_{(+)}=0.7$ and
$1.5$  are   same as those of $n_{(+)}=1.0$. Note that
for the case of $n_{(+)}=1.5$, all stellar models are  unstable if
discontinuity
gets large.

In what follows, we discuss only  stable stellar models,
i.e.
$dM/d\rho_c>0$. In our analysis, we fix the mass of a neutron star $M$
($1.2M_{\odot}$, $0.5M_{\odot}$)
because $M$ may be determined  by other  observations.  In Table
\ref{tab_property_discontinuous_1210}, Table
\ref{tab_property_discontinuous_0510} and Table
\ref{tab_property_discontinuous_1207}, we summarize the properties of
background stellar models.

% Figure 4

\subsection{Quasi-normal modes for a  stellar model without discontinuity}

In order to find  a specific property in quasi-normal modes of  a neutron
star  with
discontinuity, first we
calculate of $l=2$ in the case of a stellar model  without
discontinuity as reference.
  In this case, we have to set two parameters; the polytropic index $n$ and
polytropic
coefficient
$K$, if we fix the total  mass $M$.
We adopt the following values;
$(n,K)=(1.0,100\text{km}^{2})$, $(1.0,200\text{km}^2)$,
$(1.5,15\text{km}^{4/3})$, $(2.0,3.5\text{km})$ for $M=1.2M_{\odot}$.
We also  the case of  $M=0.5M_{\odot}$ with $(n,K)=(1.0,100\text{km}^{2})$
  to examine
the mass dependence.
The properties of the neutron
star models are summarized in Table \ref{tab_property_continuous}.

\begin{center}
\begin{figure}[htbp]
\epsfxsize = 3in
\epsffile{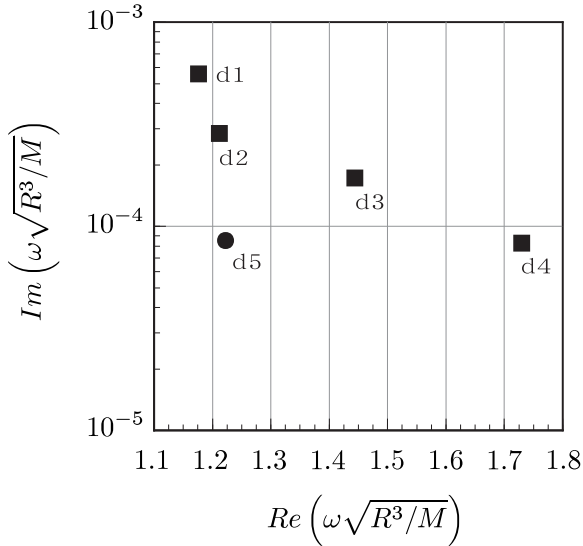}
\caption{
The quasi normal mode (f-mode) of a neutron star
  without density discontinuity.
The labels associated with each mark  correspond to the models in Table
\ref{tab_property_continuous}.
}
\label{fig_qnm_continuous}
\end{figure}
\end{center}

The quasi-normal modes of the gravitational wave emitted from such stars
are plotted in Figure \ref{fig_qnm_continuous} and listed in Table
\ref{tab_qnm_continuous}.
The plot point corresponds to the so-called f-mode of each stellar model.
  In this table we
show the mode frequency $\omega$ by two normalizations: one is
$\omega M$ and the other is $\omega(R^3/M)^{1/2}$.
Although the f-mode is determined by the radius as well as the mass,
we may not know the radius from observation. Then we can plot the data
by  $\omega M$.
In the figure, however, we only  show \ $\omega(R^3/M)^{1/2}$.
In a neutron star, we usually find many p-modes as well as f-mode,
but  we only show the lowest f-mode
here  because that we are interested in new mode which appears in low
frequency for
a star with discontinuity.

% Figure 5

 From Fig. \ref{fig_qnm_continuous}, we find that the real part
of eigenfrequency gets large as equation of state is softer.  For example,  the
frequency in the  case 'd4' is by  $47\%$ larger than that in the case 'd1'.
However, this tendency becomes reverse if we use another normalization
($\omega M$) because the radius of a neutron
star constructed by soft matter is larger than one constructed by stiff matter
(see Table \ref{tab_property_continuous}).
We also find that the imaginary part of eigenfrequency increases as
the stellar mass increases, but
the real part is insensitive to the mass.

\subsection{Quasi-normal modes for a  stellar model with discontinuity}

In the case with discontinuity, we have three unknown parameters:
$n_{(+)}$, $\rho_{\rm
dis}^{(+)}$ and $\rho_{\rm
dis}^{(-)}/\rho_{\rm
dis}^{(+)}$.
First we
show the eigenfrequency of quasi-normal modes of $l=2$ for the case of
$n_{(+)}=1.0$. We
fix the mass of a neutron star as
$M= 1.2M_{\odot}$ or
$0.5M_{\odot}$.  Now two free parameters are  $\rho_{\rm
dis}^{(+)}$ and
$\rho_{\rm
dis}^{(-)}/\rho_{\rm
dis}^{(+)}$.

\begin{center}
\begin{figure}[htbp]
\epsfxsize = 6in
\epsffile{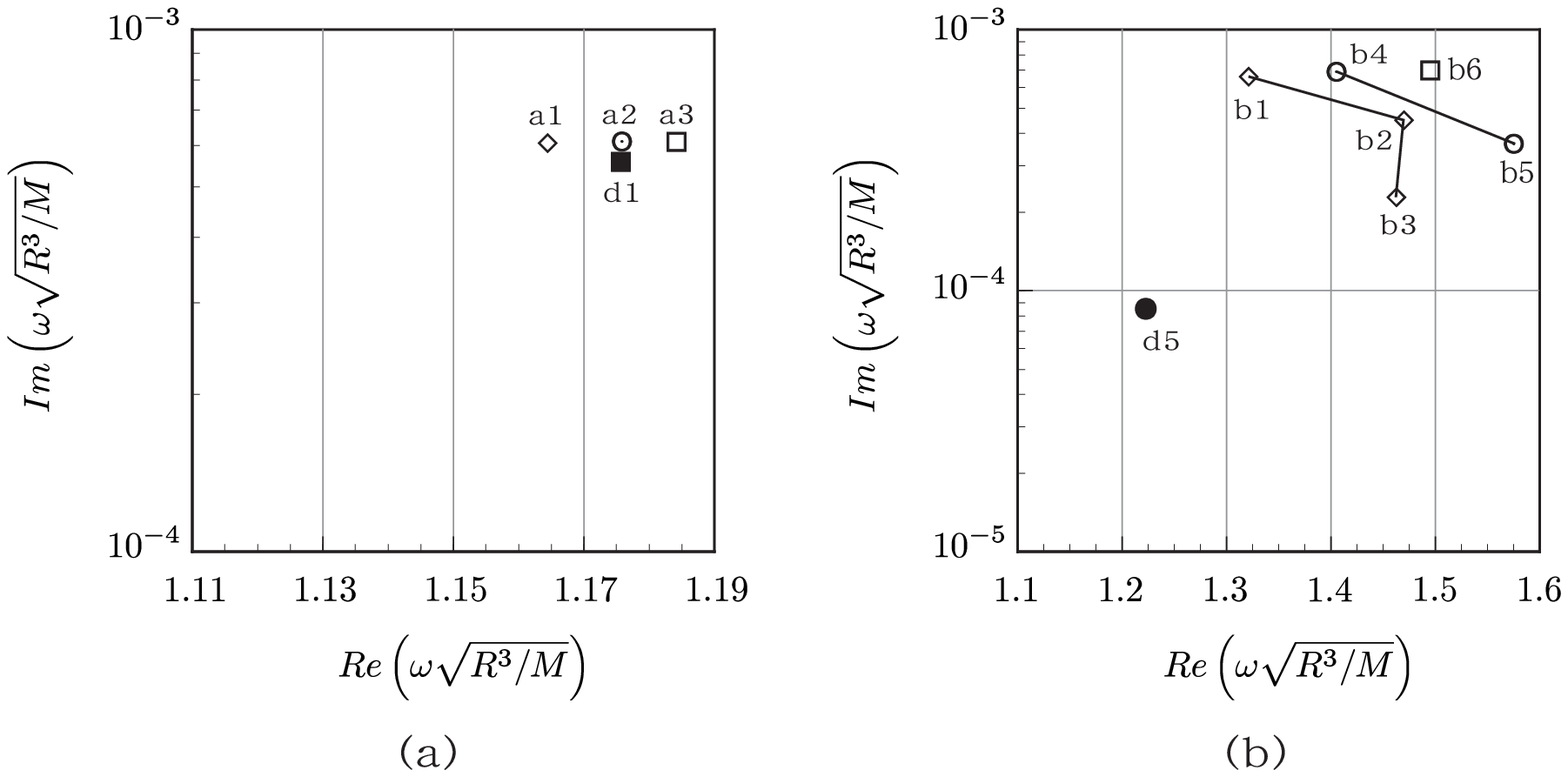}
\caption{The quasi normal mode (f-mode) of a neutron star with  density
discontinuity. We set  $n_{(+)}=1.0$. The left figure (a) shows the result  of
a star with $M=1.2M_{\odot}$, while the right one (b) corresponds to a star
with
$M=0.5M_{\odot}$. The labels associated with each mark correspond to the
models in Table
\ref{tab_property_discontinuous_1210} and in Table
\ref{tab_property_discontinuous_0510}. The data of  'd1'
and 'd5', which  are the models without discontinuity for $n=1.0$
and $K=100$\,(km$^2$), are also plotted as a reference.
}
\label{fig_qnm_f_discontinuous}
\end{figure}
\end{center}

  We show the f-mode for $M=1.2M_{\odot}$ and $0.5M_{\odot}$
in Fig. \ref{fig_qnm_f_discontinuous}.
The numerical values are given in Table
\ref{tab_qnm_f_discontinuous_120} and Table \ref{tab_qnm_f_discontinuous_50}.

% Figure 6

For f-mode, we find that the mode frequency for less massive star (0.5
$M_{\odot}$)
is more sensitive to
discontinuity  than that for massive star (1.2 $M_{\odot}$).
For example,  the difference of the model 'b5'
from  the model without discontinuity is larger than 28\% for
$0.5M_{\odot}$. It is also found that the imaginary part of
eigenfrequencies in the model with discontinuity is larger than one without
discontinuity, and this difference becomes larger as $M/R$
increases.

\begin{center}
\begin{figure}[htbp]
\epsfxsize = 6in
\epsffile{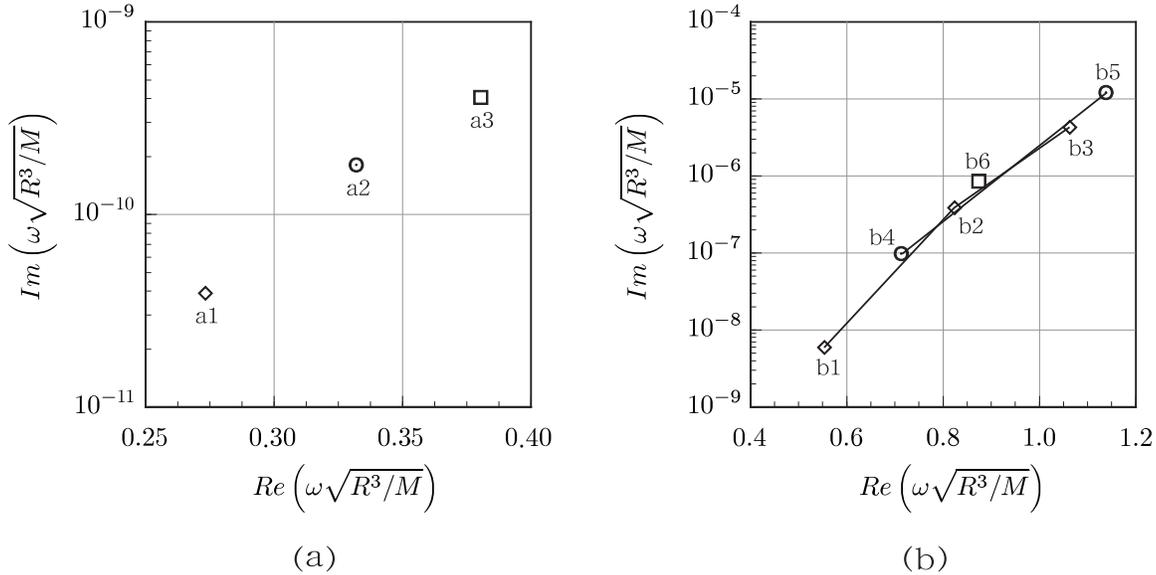}
\caption{
The quasi normal mode (g-mode) of a neutron star with  density
discontinuity. The setting is the same as Fig. 6.}
\label{fig_qnm_g_discontinuous}
\end{figure}
\end{center}

The results of g-mode for $M=1.2M_{\odot}$ and $0.5M_{\odot}$   are given
in Fig. \ref{fig_qnm_g_discontinuous} and in Table
\ref{tab_qnm_g_discontinuous_120} and Table \ref{tab_qnm_g_discontinuous_50}.
For g-mode, we also find the similar properties to the f-mode.
The mode frequency in a less  massive star is more sensitive
than that in a massive star, in particular  the imaginary part
changes by the factor
$10^2-10^4$ for the same $\rho_{\rm
dis}^{(+)}$ depending on the amount of discontinuity 
(see Fig.\ref{fig_qnm_g_discontinuous}).
Furthermore, the
existence of  g-mode is due to discontinuity, and  it turns out  that its
imaginary part
of g-mode is not always so small if the discontinuity is very large. We
conclude
that the damping rate of g-mode may not be ignored in a particular situation.

\begin{center}
\begin{figure}[htbp]
\epsfxsize = 6in
\epsffile{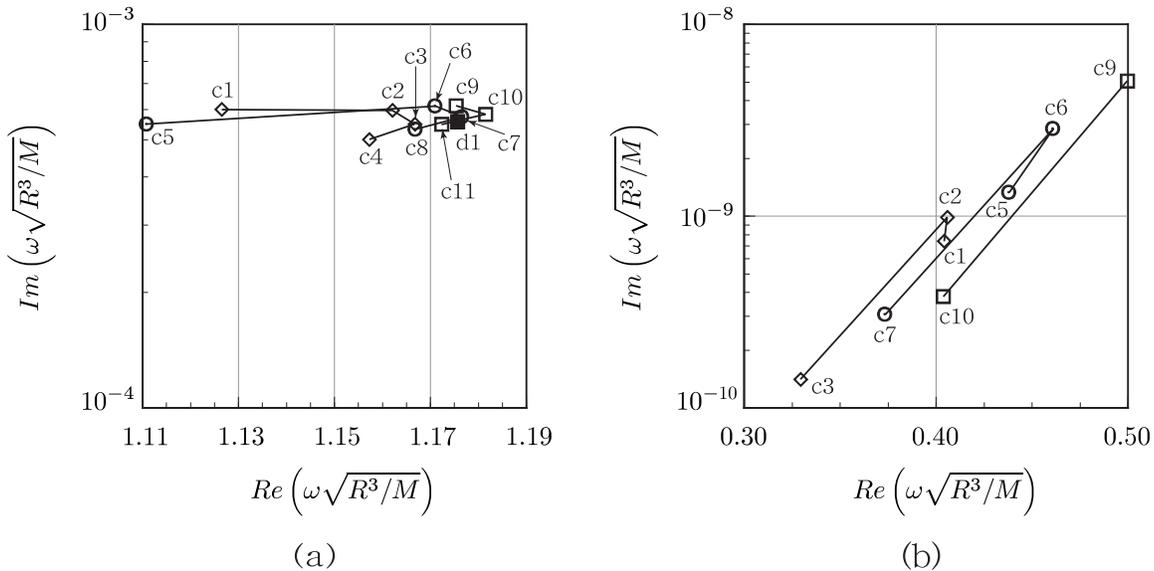}
\caption{
The quasi normal modes of a neutron star with  density
discontinuity ((a) f-mode and (b) g-mode). We set  $n_{(+)}=0.7$ and
$M=1.2M_{\odot}$.
The labels associated with each mark correspond to the models in Table
\ref{tab_property_discontinuous_1207}. The data of  'd1', which is the
model without
discontinuity for $n=1.0$  and $K=100$\,(km$^2$), are also plotted as a
reference.
}
\label{fig_qnm_discontinuous_07}
\end{figure}
\end{center}

Next we calculate the eigenfrequencies of f-mode and g-mode for
$n_{(+)}=0.7$. The
results of f-modes and g-mode are  shown  in Fig.
\ref{fig_qnm_discontinuous_07}(a) with the case of the model without
discontinuity, and
  in Fig.
\ref{fig_qnm_discontinuous_07}(b), respectively.
The numerical data  are given in
Table \ref{tab_qnm_f_discontinuous_07} and Table
\ref{tab_qnm_g_discontinuous_07}, respectively.
By comparing  the
upper panel in Fig. \ref{fig_qnm_f_discontinuous}(a), we find that the
effect of change
of equation of state is quite small.
On the other hand, the  imaginary part of g-mode
becomes by  10 times larger than the case of $n=1$ due to  the existence of
the inner
stiff matter.
  Furthermore, in this case, even if the density is continuous ($\rho_{\rm
dis}^{(-)}/\rho_{\rm
dis}^{(+)}=1.0$), one may expect the appearance of g-mode because of
discontinuity of
the equation of state ($n_{(+)}=0.7$ and $n_{(-)}=1.0$).
However, we could not find g-mode in our calculation. This may be because
the g-mode
does not exist or is too small to be found.

% Figure 7

% Figure 8

\subsection{Comparison with the Cowling approximation}

Finally, by using relativistic Cowling approximation
\cite{McDermott1983},  we shall calculate the f- and g-modes, and compare
them to our
results. In the Cowling approximation, metric  perturbation is ignored and
only stellar
oscillation is taken into account.
Hence, the eigenfrequency has only the real part. The result is given in
  Table
\ref{tab_cowling}  for $\rho_{(+)}=8.0 \times 10^{14}$\,(g/cm$^3$).
The error (the difference from our numerical value)  estimated by
  \begin{equation}
   error = \frac{\omega_{\rm Cow}-{\cal R}e(\omega)}{{\cal R}e(\omega)},
  \end{equation}
is also shown.
Here  $\omega_{\rm Cow}$ is the eigenfrequency found by the Cowling
approximation, and
$\omega_{\rm Cow}^{(g)}$  and $\omega_{\rm Cow}^{(f)}$ denote those of
g-  and f-modes, respectively.
 From Table \ref{tab_cowling},  we find that
the Cowling  approximation is very accurate for g-mode if $\rho_{(+)}$ is
at least in this region, although it is not so good for the f-mode.

\section{conclusion}
\label{sec:conclusion}

In this paper we calculate the eigenfrequencies of f-mode and g-mode to
examine the dependence of the discontinuity of the density if there exists.
We investigate with the various discontinuity, by changing $\rho_{\rm
dis}^{(+)}$
and $\rho_{\rm
dis}^{(-)}/\rho_{\rm
dis}^{(+)}$, for the case that the total mass is
fixed $1.2M_{\odot}$ for $n_{(+)}=1.0$, $0.7$ and $0.5M_{\odot}$ for
$n_{(+)}=1.0$. In consequence we find that the dependence of the
eigenfrequencies of f-mode on the discontinuity of the density is more
strongly for the case that the total mass is smaller. The effect that the
inner matter is stiff, $n_{(+)}=0.7$ and $n_{(-)}=1.0$, is quite small for
f-mode. And it is found that the imaginary part of the eigenfrequencies of
g-mode become a little large for the case that inner matter is stiff. In
general imaginary part of g-mode is quite small, but we find that in
particular situation such that an inner polytropic index is the same as an
outer one and total mass is rather small, the imaginary part of
g-mode may not be ignored.

For the considering the circumstances mentioned above, it might be possible
to judge whether the discontinuity of the density exist or not by the direct
observation of the gravitational waves, which might be practicable in
future observation, for even f-mode in the particular situation which the
total mass
is small. Of course, if the gravitational waves of g-mode
is directly observed for such stars as the small massive ones, it is found
that the discontinuity of the density exist. So we can judge the existence of
the phase transition of the density in the neutron stars by the direct
observation of the gravitational waves, though we need to collect the
detailed data of the eigenfrequencies for some characteristic mode.

Finally we find also that Cowling approximation is quite accurate for g-mode.

\acknowledgments

We would like to thank M. Saijo
for useful discussion. HS acknowledges  T. Harada for valuable comments.
This work was supported partially by JSPS Grants-in-Aid (No. 1205705)
and by the Waseda University Grant for Special Research Projects.

\newpage
\appendix

\section{the Taylor expansion at the center of a neutron star}
\label{sec:appendix_1}

Here we present the explicit perturbation equations (\ref{perturbation1})$-$
(\ref{perturbation6}) by  the expansion at the center
of a neutron  star.  $\rho$, $P$
and $\Phi$ are expanded  at $r=0$ as
  \begin{eqnarray}
   \rho(r) &=& \rho^{(0)}+\frac{1}{2}\rho^{(2)}r^{2}+O(r^{4}), \\
   P(r) &=& P^{(0)}+\frac{1}{2}P^{(2)}r^{2}+\frac{1}{4}P^{(4)}r^{4}+O(r^{6}),
           \\
   \Phi(r) &=& \Phi^{(0)}+\frac{1}{2}\Phi^{(2)}r^{2}+\frac{1}{4}\Phi^{(4)}r^{4}
     +O(r^{6}).
  \end{eqnarray}
where  the coefficients   are obtained from  Eqs.
(\ref{ephi})$-$(\ref{dpdr}) and
(\ref{gamma}) as
  \begin{eqnarray}
   \Phi^{(2)} &&= \frac{4\pi}{3}\rho^{(0)}+4\pi P^{(0)}, \\
   P^{(2)}    &&= -\left(P^{(0)}+\rho^{(0)}\right)\Phi^{(2)}, \\
   \rho^{(2)} &&= \frac{P^{(0)}+\rho^{(0)}}{P^{(0)}\gamma^{(0)}}P^{(2)}, \\
   \Phi^{(4)} &&= \frac{2\pi}{5}\rho^{(2)}+2\pi P^{(2)}
    +\frac{32\pi^{2}}{9}{\rho^{(0)}}^{2}
    +\frac{32\pi^{2}}{3}\rho^{(0)}P^{(0)}, \\
   P^{(4)}    &&= -\left(P^{(0)}+\rho^{(0)}\right)\Phi^{(4)}-\frac{1}{2}
    \left(P^{(2)}+\rho^{(2)}\right)\Phi^{(2)} .
  \end{eqnarray}
  $\gamma^{(0)}$ is the adiabatic index at $r=0$, which is defined by the
equation of state. The perturbation variables $\hat{H}_1$,
$\hat{K}$, $\hat{W}$ and $\hat{X}$ are expanded  at $r=0$
follows,
  \begin{eqnarray}
   \hat{H}_1(r) &&= {\hat{H}_{1}}^{(0)}+\frac{1}{2}{\hat{H}_{1}}^{(2)}r^{2}
    +O(r^{4}), \\
   \hat{K}(r)   &&= \hat{K}^{(0)}+\frac{1}{2}\hat{K}^{(2)}r^{2}+O(r^{4}), \\
   \hat{W}(r)   &&= \hat{W}^{(0)}+\frac{1}{2}\hat{W}^{(2)}r^{2}+O(r^{4}), \\
   \hat{X}(r)   &&= \hat{X}^{(0)}+\frac{1}{2}\hat{X}^{(2)}r^{2}+O(r^{4}),
  \end{eqnarray}
where the  leading order coefficients in these equations are fixed
by the perturbation equations (\ref{perturbation1})$-$(\ref{perturbation6}) as
  \begin{eqnarray}
   {\hat{H}^{(0)}}_{1} &&=\frac{1}{l(l+1)}\left[2l{\hat{K}^{(0)}}
     -16\pi\left({P^{(0)}}+{\rho^{(0)}}\right){\hat{W}^{(0)}}\right],
     \label{boundary1} \\
   {\hat{X}^{(0)}} &&=({P^{(0)}}+{\rho^{(0)}})e^{{\Phi^{(0)}}}
     \left[-\frac{1}{2}{\hat{K}^{(0)}}+\left(\frac{4\pi}{3}\rho^{(0)}
     +4\pi P^{(0)}-\frac{\omega^{2}}{l}e^{-2{\Phi^{(0)}}}\right)
     {\hat{W}^{(0)}}\right]. \label{boundary2}
  \end{eqnarray}
In the similar way, by using the equations (\ref{perturbation1})$-$
(\ref{perturbation6}), we find the second order coefficients   in the
expansion as
  \begin{eqnarray}
  &&\frac{l+3}{2}\hat{H_{1}}^{(2)}-\hat{K}^{(2)}
   +8\pi(P^{(0)}+\rho^{(0)})\frac{l+3}{l(l+1)}\hat{W}^{(2)}
   \nonumber \\
   &&\hspace{1cm}=\left\{\frac{4\pi}{3}(2l+3)-4\pi P^{(0)}\right\}
   \hat{H_{1}}^{(0)}-\frac{8\pi}{l}(P^{(2)}+\rho^{(2)})
   \hat{W}^{(0)}+8\pi(P^{(0)}+\rho^{(0)})Q_{1}+\frac{1}{2}Q_{0}, \\
  &&\frac{l+2}{2}\hat{K}^{(2)}-\frac{l(l+1)}{4}\hat{H_{1}}^{(2)}
   -4\pi(P^{(0)}+\rho^{(0)})\hat{W}^{(2)} \nonumber \\
   &&\hspace{1cm}= 4\pi\left[P^{(2)}+\rho^{(2)}+\frac{8\pi}{3}\rho^{(0)}
   (P^{(0)}+\rho^{(0)})\right]\hat{W}^{(0)}+\left(\frac{4\pi}{3}
   \rho^{(0)}+4\pi P^{(0)}\right)\hat{K}^{(0)}
   +\frac{1}{2}Q_{0}, \\
  &&e^{\Phi^{(0)}}(P^{(0)}+\rho^{(0)})\left[\frac{\omega^{2}}{2}
   e^{-2\Phi^{(0)}}+2\pi(P^{(0)}+\rho^{(0)})-\frac{l+2}{2}\Phi^{(2)}
   \right]\hat{W}^{(2)} \nonumber \\
   &&\hspace{1.5cm}+\frac{l+2}{2}\hat{X}^{(2)}
   +\frac{l(l+1)}{8}e^{\Phi^{(0)}}(P^{(0)}+\rho^{(0)})\hat{H_{1}}^{(2)}
    \nonumber \\
   &&\hspace{1cm}=e^{\Phi^{(0)}}(P^{(0)}+\rho^{(0)})
   \bigg[-\Phi^{(2)}\hat{K}^{(0)}
   -\frac{1}{4}Q_{0}-\frac{\omega^{2}}{2}e^{-2\Phi^{(0)}}\hat{H_{1}}^{(0)}
   -\frac{l(l+1)}{2}\Phi^{(2)}Q_{1} \nonumber \\
   &&\hspace{1.5cm}+\bigg\{(l+1)\Phi^{(4)}
   -2\pi(P^{(2)}+\rho^{(2)})-\frac{16\pi^{2}}{3}\rho^{(0)}(P^{(0)}
   +\rho^{(0)})-\frac{4\pi}{3}\rho^{(0)}\Phi^{(2)} \nonumber \\
   &&\hspace{1.5cm}+\Phi^{(4)}
   +\omega^2 e^{-2\Phi^{(0)}}\left(\Phi^{(2)}-\frac{4\pi}{3}\rho^{(0)}
   \right)\bigg\}\hat{W}^{(0)}\bigg]+\frac{l}{2}\left(\Phi^{(2)}
   +\frac{P^{(2)}+\rho^{(2)}}{P^{(0)}+\rho^{(0)}}\right)
   \hat{X}^{(0)}, \\
  &&-\frac{1}{4}(P^{(0)}+\rho^{(0)})\hat{K}^{(2)}
   -\frac{1}{2}\left[P^{(2)}+\frac{l+3}{l(l+1)}\omega^{2} e^{-2\Phi^{(0)}}
   (P^{(0)}+\rho^{(0)})\right]\hat{W}^{(2)}
   -\frac{1}{2}e^{-\Phi^{(0)}}\hat{X}^{(2)} \nonumber \\
   &&\hspace{1cm}=-\frac{1}{2} e^{-\Phi^{(0)}}\Phi^{(2)}\hat{X}^{(0)}
   +\frac{1}{4}(P^{(2)}+\rho^{(2)})\hat{K}^{(0)}
   +\frac{1}{4}(P^{(0)}+\rho^{(0)})Q_{0}
   \nonumber \\
   &&\hspace{1.5cm} +\left[P^{(4)}-\frac{4\pi}{3}\rho^{(0)}P^{(2)}
   +\frac{\omega^2}{l}e^{-2\Phi^{(0)}}\left\{\frac{1}{2}(P^{(2)}+\rho^{(2)})
   -(P^{(0)}+\rho^{(0)})\Phi^{(2)}\right\}\right]\hat{W}^{(0)} \nonumber \\
   &&\hspace{1.5cm}-\frac{1}{2}\omega^2 e^{-2\Phi^{(0)}}(P^{(0)}
   +\rho^{(0)})Q_{1},
  \end{eqnarray}
where  $Q_0$ and $Q_1$ are  defined by
  \begin{eqnarray}
  &&Q_{0}\equiv
   \frac{4}{(l+2)(l-1)}\biggl[-8\pi e^{-\Phi^{(0)}}\hat{X}^{(0)}
   -\left(\omega^{2}
   e^{-2\Phi^{(0)}}+\frac{8\pi}{3}\rho^{(0)}\right)\hat{K}^{(0)}
    \nonumber \\
   &&\hspace{3.5cm}+\left\{\omega^{2}e^{-2\Phi^{(0)}}-l(l+1)
   \left(\frac{2\pi}{3}\rho^{(0)}
   +2\pi P^{(0)}\right)\right\}\hat{H_{1}}^{(0)}\biggr], \\
  &&Q_{1}\equiv
   \frac{2}{l(l+1)}\left[\frac{1}{\gamma^{(0)}P^{(0)}}e^{-\Phi^{(0)}}
   \hat{X}^{(0)}-\frac{3}{2}\hat{K}^{(0)}
   +\frac{4\pi(l+1)}{3}\rho^{(0)}\hat{W}^{(0)}\right].
  \end{eqnarray}
$\hat{H}$ and $\hat{V}$ are also expanded  at $r=0$ as
  \begin{eqnarray}
   \hat{H}(r)  &&= \hat{H}^{(0)}+\frac{1}{2}\hat{H}^{(2)}r^{2}+O(r^{4}), \\
   \hat{V}(r)  &&= \hat{V}^{(0)}+\frac{1}{2}\hat{V}^{(2)}r^{2}+O(r^{4}),
  \end{eqnarray}
where the coefficients are given by
  \begin{eqnarray}
   \hat{H}^{(0)} &&= \hat{K}^{(0)}, \\
   \hat{V}^{(0)} &&= -\frac{1}{l}\hat{W}^{(0)}, \\
   \hat{H}^{(2)} &&=Q_{0}+\hat{K}^{(2)}, \\
   \hat{V}^{(2)} &&=Q_{1}-\frac{l+3}{l(l+1)}\hat{W}^{(2)}.
  \end{eqnarray}

\newpage
\twocolumn
\section{How to  calculate the quasi normal modes}
\label{sec:appendix_2}

To calculate the quasi normal modes of a neutron star, we follow the method
developed by
  Leins, Nollert and Soffel \cite{Leins_Nollert}.
First, we change the variables from
Zerilli's one ($Z$) to the Regge-Wheeler's one ($Q$), because the
effective potential in the  Regge-Wheeler equation is very simple. The
relation between
those two variables given by\cite{Chandrasekhar_Math}
  \begin{eqnarray}
   \left(\kappa+2i\omega\beta\right)Z &=& \left(\kappa+2\beta^{2}f(r)\right)Q
     +2\beta\frac{dQ}{dr_{*}}, \label{hennkann1} \\
   \left(\kappa-2i\omega\beta\right)Q &=& \left(\kappa+2\beta^{2}f(r)\right)Z
     -2\beta\frac{dZ}{dr_{*}},  \label{hennkann2}
  \end{eqnarray}
where
  \begin{eqnarray}
    \beta  &\equiv& 6M, \\
    \kappa &\equiv& 4\lambda(\lambda+1), \\
    f(r)   &\equiv& \frac{r-2M}{2r^{2}(\lambda r+3M)}.
  \end{eqnarray}
To find the solution of the Regge-Wheeler with
appropriate  boundary condition at infinity, we expand the variable
$Q$ at a finite radius $r=r_{a}$ as
  \begin{equation}
   Q(r) = \left(\frac{r}{2M}-1\right)^{-2i M\omega}e^{-i\omega r}
     \sum_{n=0}^{\infty}a_{n}\left(1-\frac{r_{a}}{r}\right)^{n}.
  \end{equation}
By substituting this into the  Regge-Wheeler equation, we obtain a four-term
recurrence relation for $a_{n}$ ($n\ge 2$) ;
  \begin{equation}
   \alpha_{n}a_{n+1}+\beta_{n}a_{n}+\gamma_{n}a_{n-1}+\delta_{n}a_{n-2}=0,
  \end{equation}
where
  \begin{eqnarray}
   \alpha_{n} &=&\left(1-\frac{2M}{r_{a}}\right)n(n+1), \\
   \beta_{n}  &=&-2\left\{i\omega r_{a}
     +\left(1-\frac{3M}{r_{a}}\right)n\right\}n, \\
   \gamma_{n} &=&\left(1-\frac{6M}{r_{a}}\right)n(n-1)+\frac{6M}{r_{a}}-l(l+1),
   \\
   \delta_{n} &=&\frac{2M}{r_{a}}(n-3)(n+1).
  \end{eqnarray}
The coefficients  $a_{-1}$, $a_0$,  and $a_1$ are given  by the value of
$Q$ and $dQ/dr$
at $r=r_a$ as
  \begin{eqnarray}
   &&a_{-1} = 0, \\
   &&a_{0}  = \frac{Q(r_{a})}{\chi(r_{a})}, \\
   &&a_{1}  = \frac{r_{a}}{\chi(r_{a})}\left[\frac{dQ(r_{a})}{dr}
      +\frac{i\omega r_{a}}{r_{a}-2M}Q(r_{a})\right],
  \end{eqnarray}
where
  \begin{equation}
   \chi(r)=\left(\frac{r}{2M}-1\right)^{-2i M\omega}e^{-i \omega r}.
  \end{equation}
The convergence condition ($|a_{n+1}/a_{n}|<1$) is now
$r_{a}>4M$\cite{Benhar_Berti}.
If $R>4M$, we set
$r_a=R$, while  if $R\le 4M$, we integrate the Regge-Wheeler equation from
$r=R$ to
$r=r_a (>4M)$ to find the values of
$Q(r_a)$ and $dQ(r_a)/dr$. Furthermore, defining
new coefficients
$\hat{\alpha}_{n}$, $\hat{\beta}_{n}$, and $\hat{\gamma}_{n}$ as
  \begin{equation}
   \hat{\alpha}_{1}=\alpha_{1},\,\, \hat{\beta}_{1}=\beta_{1},\,\,
  \hat{\gamma}_{1}=\gamma_{1},
  \end{equation}
and for $n\ge 2$
\begin{eqnarray}
   \hat{\alpha}_{n}&=&\alpha_{n}, \\
   \hat{\beta}_{n}&=&\beta_{n}-\frac{\hat{\alpha}_{n-1}
      \delta_{n}}{\hat{\gamma}_{n-1}}, \\
   \hat{\gamma}_{n}&=&\gamma_{n}-\frac{\hat{\beta}_{n-1}
      \delta_{n}}{\hat{\gamma}_{n-1}}, \\
   \hat{\delta}_{n}&=&0,
  \end{eqnarray}
the above four-term recurrence relation
is reduced  into a three-term recurrence relation
  \begin{equation}
   \hat{\alpha}_{n}a_{n+1}+\hat{\beta}_{n}a_{n}+\hat{\gamma}_{n}a_{n-1}=0.
  \end{equation}
Using this three-term recurrence relation, we find the relation between
$\hat{\alpha}_{n}$, $\hat{\beta}_{n}$ and $\hat{\gamma}_{n}$ as the form of
continued
fraction:
  \begin{equation}
   \frac{a_{1}}{a_{0}}=\frac{-\hat{\gamma}_{1}}{\hat{\beta}_{1}-}
      \frac{\hat{\alpha}_{1}\hat{\gamma}_{2}}{\hat{\beta}_{2}-}
      \frac{\hat{\alpha}_{2}\hat{\gamma}_{3}}{\hat{\beta}_{3}-}\cdots,
      \label{conti}
  \end{equation}
which can be rewritten as
  \begin{equation}
   0=\hat{\beta}_{0}-\frac{\hat{\alpha}_{0}
      \hat{\gamma}_{1}}{\hat{\beta}_{1}-}\frac{\hat{\alpha}_{1}
      \hat{\gamma}_{2}}{\hat{\beta}_{2}-}\frac{\hat{\alpha}_{2}
      \hat{\gamma}_{3}}{\hat{\beta}_{3}-}\cdots
\equiv
f(\omega) ,
\label{tennkai1}
  \end{equation}
where $\hat{\beta}_{0}\equiv a_{1}/a_{0}$, $\hat{\alpha}_{0}\equiv-1$.
The eigenfrequency $\omega$ of  a quasi normal mode is obtained by solving
the equation
$f(\omega)=0$.

%\newpage
% Reference

\newpage

\onecolumn

% Table 1
\begin{table}
\caption{The properties of a neutron star  constructed by the
  equation of state with discontinuity. We set  $M=1.2M_{\odot}$ and
$n_{(+)}=1.0$.}
\label{tab_property_discontinuous_1210}
  \begin{center}
    \begin{tabular}{c|c|c|c|c|c|c|c}
      label & $\rho_{\rm
dis}^{(+)}\,\,(\text{g/cm}^3)$ &
              $\rho_{\rm
dis}^{(-)}/\rho_{\rm
dis}^{(+)}$ &
              $\rho_c\,\,(\text{g/cm}^3) $ & $R\,\,(\text{km})$ &
              $R_{\rm
dis}/R\,\,(\%)$ & $M_{\rm
dis}/M\,\,(\%)$ & $M/R$\\  \hline
      a1  & $8.00\times 10^{14}$ & $0.9$   & $5.017\times 10^{15}$
          & $7.577$               & $81.42$ & $89.23$ & $0.2339$ \\
      a2  & $1.20\times 10^{15}$ & $0.9$   & $4.716\times 10^{15}$
          & $7.800$               & $73.13$ & $78.20$ & $0.2272$ \\
      a3  & $1.60\times 10^{15}$ & $0.9$   & $4.278\times 10^{15}$
          & $8.112$               & $64.02$ & $62.90$ & $0.2184$
    \end{tabular}
  \end{center}
\end{table}

% Table 2
\begin{table}
\caption{The properties of a neutron star  constructed by the
  equation of state with discontinuity. We set  $M=0.5M_{\odot}$ and
$n_{(+)}=1.0$.}
\label{tab_property_discontinuous_0510}
  \begin{center}
    \begin{tabular}{c|c|c|c|c|c|c|c}
      label & $\rho_{\rm
dis}^{(+)}\,\,(\text{g/cm}^3)$ &
              $\rho_{\rm
dis}^{(-)}/\rho_{\rm
dis}^{(+)}$ &
              $\rho_c\,\,(\text{g/cm}^3) $ & $R\,\,(\text{km})$ &
              $R_{\rm
dis}/R\,\,(\%)$ & $M_{\rm
dis}/M\,\,(\%)$ & $M/R$\\  \hline
      b1  & $8.00\times 10^{14}$ & $0.4$   & $1.759\times 10^{16}$
          & $3.925$               & $86.99$ & $98.76$ & $0.1881$  \\
      b2  & $8.00\times 10^{14}$ & $0.5$   & $5.800\times 10^{15}$
          & $6.050$               & $73.65$ & $91.10$ & $0.1220$  \\
      b3  & $8.00\times 10^{14}$ & $0.6$   & $2.484\times 10^{15}$
          & $8.465$               & $56.76$ & $65.25$ & $0.0872$1 \\
      b4  & $1.20\times 10^{15}$ & $0.4$   & $1.733\times 10^{16}$
          & $4.137$               & $80.71$ & $97.16$ & $0.1785$  \\
      b5  & $1.20\times 10^{15}$ & $0.5$   & $5.127\times 10^{15}$
          & $6.985$               & $59.38$ & $76.64$ & $0.1057$  \\
      b6  & $1.60\times 10^{15}$ & $0.4$   & $1.680\times 10^{16}$
          & $4.407$               & $74.28$ & $94.69$ & $0.1675$
    \end{tabular}
  \end{center}
\end{table}

% Table 3
\begin{table}
\caption{
The properties of a neutron star  constructed by the
  equation of state with discontinuity. We set  $M=1.2M_{\odot}$ and
$n_{(+)}=0.7$.
}
\label{tab_property_discontinuous_1207}
  \begin{center}
    \begin{tabular}{c|c|c|c|c|c|c|c}
      label & $\rho_{\rm
dis}^{(+)}\,\,(\text{g/cm}^3)$ &
              $\rho_{\rm
dis}^{(-)}/\rho_g^{(+)}$ &
              $\rho_c\,\,(\text{g/cm}^3) $ & $R\,\,(\text{km})$ &
              $R_{\rm
dis}/R\,\,(\%)$ & $M_{\rm
dis}/M\,\,(\%)$ & $M/R$\\  \hline
      c1  & $8.00\times 10^{14}$ & $0.7$    & $4.237 \times 10^{15}$
          & $7.351$               & $85.64$  & $93.61$ & $0.2410$ \\
      c2  & $8.00\times 10^{14}$ & $0.8$    & $2.787 \times 10^{15}$
          & $8.488$               & $79.57$  & $85.53$ & $0.2088$ \\
      c3  & $8.00\times 10^{14}$ & $0.9$    & $2.029 \times 10^{15}$
          & $9.387$               & $73.47$  & $74.03$ & $0.1888$ \\
      c4  & $8.00\times 10^{14}$ & $1.0$    & $1.580 \times 10^{15}$
          & $10.04$               & $67.38$  & $60.31$ & $0.1765$ \\
      c5  & $1.20\times 10^{15}$ & $0.7$    & $5.924 \times 10^{15}$
          & $6.724$               & $83.11$  & $91.86$ & $0.2635$ \\
      c6  & $1.20\times 10^{15}$ & $0.8$    & $3.483 \times 10^{15}$
          & $8.090$               & $73.88$  & $78.54$ & $0.2190$ \\
      c7  & $1.20\times 10^{15}$ & $0.9$    & $2.449 \times 10^{15}$
          & $9.046$               & $65.08$  & $60.68$ & $0.1959$ \\
      c8  & $1.20\times 10^{15}$ & $1.0$    & $1.903 \times 10^{15}$
          & $9.634$               & $56.62$  & $42.14$ & $0.1839$ \\
      c9  & $1.60\times 10^{15}$ & $0.8$    & $4.068 \times 10^{15}$
          & $7.862$               & $68.95$  & $71.75$ & $0.2254$ \\
      c10 & $1.60\times 10^{15}$ & $0.9$    & $2.740 \times 10^{15}$
          & $8.912$               & $56.79$  & $46.69$ & $0.1988$ \\
      c11 & $1.60\times 10^{15}$ & $1.0$    & $2.146 \times 10^{15}$
          & $9.421$               & $45.71$  & $25.46$ & $0.1881$
    \end{tabular}
  \end{center}
\end{table}

% Table 4
\begin{table}
\caption{The properties of a neutron star  constructed by the
equation of state without  discontinuity.
The total mass $M$ is fixed $1.2M_{\odot}$ or $0.5M_{\odot}$.}
\label{tab_property_continuous}
  \begin{center}
    \begin{tabular}{c|c|c|c|c|c|c}
      label & Mass\,\,($M_{\odot}$) & $n$  
               & $K$ & $\rho_c\,\,(\text{g/cm}^3) $
            & $R\,\,(\text{km})$    & $M/R$     \\  \hline
      d1 & $1.2$ & $1.0$ & $100$\,(km$^2$)    & $2.427 \times 10^{15}$
         & $9.296$                & $0.1906$  \\
      d2 & $1.2$ & $1.0$ & $200$\,(km$^2$)    & $5.507 \times 10^{14}$
         & $15.10$                & $0.1173$  \\
      d3 & $1.2$ & $1.5$ & $15$\,(km$^{4/3}$) & $2.691 \times 10^{14}$
         & $23.49$                & $0.07542$ \\
      d4 & $1.2$ & $2.0$ & $3.5$\,(km)        & $1.153 \times 10^{14}$
         & $38.79$                & $0.04568$ \\
      d5 & $0.5$ & $1.0$ & $100$\,(km$^2$)    & $5.067  \times 10^{14}$
         & $11.57$                & $0.06384$
     \end{tabular}
  \end{center}
\end{table}

% Table 5
\begin{table}
\caption{The numerical data of quasi normal mode (f-mode) of the models in
Table\ref{tab_property_continuous}.
These data are plotted in Fig.\ref{fig_qnm_continuous} .}
\label{tab_qnm_continuous}
  \begin{center}
    \begin{tabular}{c|c|c|c|c}
      label & ${\cal R}e\left(\omega M\right)$ &
              ${\cal I}m\left(\omega M\right)$ &
              ${\cal R}e\left(\omega\sqrt{R^3/M}\right)$ &
              ${\cal I}m\left(\omega\sqrt{R^3/M}\right)$ \\  \hline
      d1 & $0.09784$  & $4.643 \times 10^{-5}$
         & $1.176$    & $5.579 \times 10^{-4}$ \\
      d2 & $0.04869$  & $1.145 \times 10^{-5}$
         & $1.211$    & $2.849 \times 10^{-4}$ \\
      d3 & $0.02990$  & $3.572 \times 10^{-6}$
         & $1.443$    & $1.725 \times 10^{-4}$ \\
      d4 & $0.01688$  & $8.075 \times 10^{-7}$
         & $1.729$    & $8.270 \times 10^{-5}$ \\
      d5 & $0.01974$  & $1.374 \times 10^{-6}$
         & $1.224$    & $8.517 \times 10^{-5}$
    \end{tabular}
  \end{center}
\end{table}

% Table 6
\begin{table}
\caption{
The numerical data of quasi normal mode (f-mode) of the models in
Table\ref{tab_property_discontinuous_1210} ($M=1.2M_{\odot}$ and
$n_{(+)}=1.0$).
These data are plotted in Fig.\ref{fig_qnm_f_discontinuous} (a).}
\label{tab_qnm_f_discontinuous_120}
  \begin{center}
    \begin{tabular}{c|c|c|c|c}
      label & ${\cal R}e\left(\omega M\right)$ &
              ${\cal I}m\left(\omega M\right)$ &
              ${\cal R}e\left(\omega\sqrt{R^3/M}\right)$ &
              ${\cal I}m\left(\omega\sqrt{R^3/M}\right)$ \\  \hline
      a1  & $0.1317$   & $6.856\times 10^{-5}$
          & $1.164$    & $6.062\times 10^{-4}$ \\
      a2  & $0.1273$   & $6.604\times 10^{-5}$
          & $1.176$    & $6.098\times 10^{-4}$ \\
      a3  & $0.1209$   & $6.221\times 10^{-5}$
          & $1.184$    & $6.093\times 10^{-4}$
    \end{tabular}
  \end{center}
\end{table}

% Table 7
\begin{table}
\caption{
The numerical data of quasi normal mode (f-mode) of the models in
Table\ref{tab_property_discontinuous_0510} ($M=0.5M_{\odot}$ and
$n_{(+)}=1.0$).
These data are plotted in Fig.\ref{fig_qnm_f_discontinuous} (b).
}
\label{tab_qnm_f_discontinuous_50}
  \begin{center}
    \begin{tabular}{c|c|c|c|c}
      label & ${\cal R}e\left(\omega M\right)$ &
              ${\cal I}m\left(\omega M\right)$ &
              ${\cal R}e\left(\omega\sqrt{R^3/M}\right)$ &
              ${\cal I}m\left(\omega\sqrt{R^3/M}\right)$ \\  \hline
      b1  & $0.1078$   & $5.384\times 10^{-5}$
          & $1.321$    & $6.600\times 10^{-4}$ \\
      b2  & $0.06266$  & $1.916\times 10^{-5}$
          & $1.470$    & $4.495\times 10^{-4}$ \\
      b3  & $0.03767$  & $5.867\times 10^{-6}$
          & $1.462$    & $2.278\times 10^{-4}$ \\
      b4  & $0.1060$   & $5.185\times 10^{-5}$
          & $1.406$    & $6.878\times 10^{-4}$ \\
      b5  & $0.05416$  & $1.253\times 10^{-5}$
          & $1.576$    & $3.645\times 10^{-4}$ \\
      b6  & $0.1025$   & $4.780\times 10^{-5}$
          & $1.495$    & $6.971\times 10^{-4}$
    \end{tabular}
  \end{center}
\end{table}

% Table 8
\begin{table}
\caption{
The numerical data of quasi normal mode (g-mode) of the models in
Table\ref{tab_property_discontinuous_1210} ($M=1.2M_{\odot}$ and
$n_{(+)}=1.0$).
These data are plotted in Fig.\ref{fig_qnm_g_discontinuous} (a).
}
\label{tab_qnm_g_discontinuous_120}
  \begin{center}
    \begin{tabular}{c|c|c|c|c}
      label & ${\cal R}e\left(\omega M\right)$ &
              ${\cal I}m\left(\omega M\right)$ &
              ${\cal R}e\left(\omega\sqrt{R^3/M}\right)$ &
              ${\cal I}m\left(\omega\sqrt{R^3/M}\right)$ \\  \hline
      a1  & $3.090\times 10^{-2}$ & $4.410\times 10^{-12}$
          & $0.2732$               & $3.899\times 10^{-11}$ \\
      a2  & $3.598\times 10^{-2}$ & $1.955\times 10^{-11}$
          & $0.3322$               & $1.806\times 10^{-10}$ \\
      a3  & $3.884\times 10^{-2}$ & $4.137\times 10^{-11}$
          & $0.3804$               & $4.052\times 10^{-10}$ \\
    \end{tabular}
  \end{center}
\end{table}

% Table 9
\begin{table}
\caption{
The numerical data of quasi normal mode (g-mode) of the models in
Table\ref{tab_property_discontinuous_0510} ($M=0.5M_{\odot}$ and
$n_{(+)}=1.0$).
These data are plotted in Fig.\ref{fig_qnm_g_discontinuous} (b).
}
\label{tab_qnm_g_discontinuous_50}
  \begin{center}
    \begin{tabular}{c|c|c|c|c}
      label & ${\cal R}e\left(\omega M\right)$ &
              ${\cal I}m\left(\omega M\right)$ &
              ${\cal R}e\left(\omega\sqrt{R^3/M}\right)$ &
              ${\cal I}m\left(\omega\sqrt{R^3/M}\right)$ \\  \hline
      b1  & $4.514\times 10^{-2}$ & $4.849\times 10^{-10}$
          & $0.5532$                        & $5.944\times 10^{-9}$ \\
      b2  & $3.510\times 10^{-2}$ & $1.650\times 10^{-8}$
          & $0.8234$                        & $3.870\times 10^{-7}$ \\
      b3  & $2.737\times 10^{-2}$ & $1.104\times 10^{-7}$
          & $1.062$                          & $4.287\times 10^{-6}$ \\
      b4  & $5.387\times 10^{-2}$ & $7.374\times 10^{-9}$
          & $0.7145$                        & $9.781\times 10^{-8}$ \\
      b5  & $3.916\times 10^{-2}$ & $4.150\times 10^{-7}$
          & $1.139$                          & $1.208\times 10^{-5}$ \\
      b6  & $5.992\times 10^{-2}$ & $5.865\times 10^{-8}$
          & $0.8738$                        & $8.553\times 10^{-7}$
    \end{tabular}
  \end{center}
\end{table}

% Table 10
\begin{table}
\caption{
The numerical data of quasi normal mode (f-mode) of the models in
Table\ref{tab_property_discontinuous_1207} ($M=1.2M_{\odot}$ and
$n_{(+)}=0.7$).
These data are plotted in Fig.\ref{fig_qnm_discontinuous_07} (a).
}
\label{tab_qnm_f_discontinuous_07}
  \begin{center}
    \begin{tabular}{c|c|c|c|c}
      label & ${\cal R}e\left(\omega M\right)$ &
              ${\cal I}m\left(\omega M\right)$ &
              ${\cal R}e\left(\omega\sqrt{R^3/M}\right)$ &
              ${\cal I}m\left(\omega\sqrt{R^3/M}\right)$ \\  \hline
      c1  & $0.1333$  & $7.102 \times 10^{-5}$
          & $1.126$   & $6.001 \times 10^{-4}$ \\
      c2  & $0.1108$  & $5.692 \times 10^{-5}$
          & $1.162$   & $5.967 \times 10^{-4}$ \\
      c3  & $0.09569$ & $4.503 \times 10^{-5}$
          & $1.167$   & $5.490 \times 10^{-4}$ \\
      c4  & $0.08579$ & $3.713 \times 10^{-5}$
          & $1.157$   & $5.008 \times 10^{-4}$ \\
      c5  & $0.1503$  & $7.432 \times 10^{-5}$
          & $1.111$   & $5.494 \times 10^{-4}$ \\
      c6  & $0.1200$  & $6.278 \times 10^{-5}$
          & $1.171$   & $6.125 \times 10^{-4}$ \\
      c7  & $0.1020$  & $4.975 \times 10^{-5}$
          & $1.177$   & $5.739 \times 10^{-4}$ \\
      c8  & $0.09205$ & $4.201 \times 10^{-5}$
          & $1.167$   & $5.326 \times 10^{-4}$ \\
      c9  & $0.1258$  & $6.560 \times 10^{-5}$
          & $1.175$   & $6.130 \times 10^{-4}$ \\
      c10 & $0.1047$  & $5.163 \times 10^{-5}$
          & $1.181$   & $5.824 \times 10^{-4}$ \\
      c11 & $0.09564$ & $4.476 \times 10^{-5}$
          & $1.172$   & $5.487 \times 10^{-4}$
    \end{tabular}
  \end{center}
\end{table}

% Table 11
\begin{table}
\caption{
The numerical data of quasi normal mode (g-mode) of the models in
Table\ref{tab_property_discontinuous_1207} ($M=1.2M_{\odot}$ and
$n_{(+)}=0.7$).
These data are plotted in Fig.\ref{fig_qnm_discontinuous_07} (b).
}
\label{tab_qnm_g_discontinuous_07}
  \begin{center}
    \begin{tabular}{c|c|c|c|c}
      label & ${\cal R}e\left(\omega M\right)$ &
              ${\cal I}m\left(\omega M\right)$ &
              ${\cal R}e\left(\omega\sqrt{R^3/M}\right)$ &
              ${\cal I}m\left(\omega\sqrt{R^3/M}\right)$ \\  \hline
      c1  & $4.782 \times 10^{-2}$ & $8.763 \times 10^{-11}$
          & $0.4040$                         & $7.404 \times 10^{-10}$ \\
      c2  & $3.870 \times 10^{-2}$ & $9.396 \times 10^{-11}$
          & $0.4057$                         & $9.851 \times 10^{-10}$ \\
      c3  & $2.701 \times 10^{-2}$ & $1.154 \times 10^{-11}$
          & $0.3293$                         & $1.407 \times 10^{-10}$ \\
      c5  & $5.926 \times 10^{-2}$ & $1.799 \times 10^{-10}$
          & $0.4381$                         & $1.330 \times 10^{-9}$ \\
      c6  & $4.725 \times 10^{-2}$ & $2.931 \times 10^{-10}$
          & $0.4610$                         & $2.859 \times 10^{-9}$ \\
      c7  & $3.238 \times 10^{-2}$ & $2.666 \times 10^{-11}$
          & $0.3735$                         & $3.075 \times 10^{-10}$ \\
      c9  & $5.346 \times 10^{-2}$ & $5.412 \times 10^{-10}$
          & $0.4996$                         & $5.058 \times 10^{-9}$ \\
      c10 & $3.578 \times 10^{-2}$ & $3.374 \times 10^{-11}$
          & $0.4036$                         & $3.805 \times 10^{-10}$
    \end{tabular}
  \end{center}
\end{table}

% Table 12
\begin{table}
\caption{Comparison with the results of Cowling approximation.
The labels in the table corresponds to each model.
The data in this Table is only for $\rho_{(+)}=8.0\times 10^{14}$\,(g/cm$^3$).
}
\label{tab_cowling}
  \begin{center}
    \begin{tabular}{c|c|c|c|c}
      label & $\omega_{\rm Cow}^{(g)}\sqrt{R^3/M}$ &
              $error^{(g)}\,(\%)$ &
              $\omega_{\rm Cow}^{(f)}\sqrt{R^3/M}$ &
              $error^{(f)}\,(\%)$ \\  \hline
      a1  & $0.2689$ & $-1.570$
          & $1.304$  & $12.01$ \\
      b1  & $0.5533$ & $0.01808$
          & $1.520$  & $15.10$ \\
      b2  & $0.8329$ & $1.161$
          & $1.733$  & $17.94$ \\
      b3  & $1.107$  & $4.216$
          & $1.717$  & $17.41$ \\
      c1  & $0.4015$ & $-0.6188$
          & $1.291$  & $14.63$ \\
      c2  & $0.4016$ & $-1.003$
          & $1.355$  & $16.57$ \\
      c3  & $0.3232$ & $-1.837$
          & $1.379$  & $18.21$ \\
      c4  & $---$    & $---$
          & $1.388$  & $19.97$
    \end{tabular}
  \end{center}
\end{table}

\end{document}